\documentclass[preprint,pre]{revtex4}
\usepackage{amsfonts}
\usepackage{amssymb}
\usepackage{color}  
\usepackage{graphicx}

\newcommand{\Br}{{\bf r}}

\newcommand{\BR}{{\bf R}}

\newcommand\s{{\bf \hat s}}

\begin{document}

\title{Single-Scattering Optical Tomography} 

\author{Lucia Florescu}

\affiliation{Department of Bioengineering, University of Pennsylvania,
  Philadelphia, PA 19104}

\author{John C. Schotland}

\affiliation{Department of Bioengineering and Graduate Program in Applied Mathematics and Computational Science, University of Pennsylvania,
Philadelphia, PA 19104}

\author{Vadim A. Markel}

\affiliation{Department of Radiology, University of Pennsylvania,
  Philadelphia, PA 19104}

\begin{abstract}
We consider the problem of  optical tomographic imaging in the
mesoscopic regime where the photon mean free path is of order of the
system size. Within the accuracy of the single-scattering
approximation to the radiative transport equation, we show that it is
possible to recover the extinction coefficient of an inhomogeneous
medium from angularly-resolved measurements. Applications to
biomedical imaging are described and illustrated with numerical
simulations.
\end{abstract}

\date{\today} 

\maketitle

\section{Introduction}
\label{sec:intro}

There has been considerable recent interest in the development of
experimental methods for three-dimensional optical imaging of
biological systems. Applications range from imaging of optically-thin
(by which we mean nearly transparent) cellular and sub-cellular
structures to optically-thick systems at the whole organ level in
which multiple scattering of light occurs. In optically-thin systems,
confocal microscopy~\cite{Wilson} and optical coherence
microscopy~\cite{Izatt_1994} can be used to generate three-dimensional
images by optical sectioning. Alternatively, computed imaging methods
such as optical projection tomography (OPT)~\cite{Sharpe_2002} or
interferometric synthetic aperture microscopy
(ISAM)~\cite{Ralston_2006,Ralston_2007} may be employed to reconstruct
three-dimensional images by inversion of a suitable integral
equation. In the case of OPT, the effects of scattering are ignored
and geometrical optics is used to describe the propagation of
light. The sample structure is then recovered by inversion of a Radon
transform which relates the extinction coefficient of the sample to
the measured intensity of the optical field. In the case of ISAM, the
effects of scattering are accounted for within the accuracy of the
first Born approximation to the wave equation. An inverse scattering
problem is then solved to recover the susceptibility of the sample
from interferometric measurements of the cross-correlation function of
the optical field.

In optically-thick systems, multiple scattering of the illuminating
field creates a fundamental obstruction to image formation. If the
medium is macroscopically large and weakly absorbing, only diffuse
light is transmitted. By making use of the diffusion approximation
(DA) to the radiative transport equation (RTE) and solving an
appropriate inverse problem, the aforementioned difficulty may, to
some extent, be overcome. This approach forms the basis of diffusion
tomography which can be used to reconstruct images with sub-centimeter
resolution of highly-scattering media such as the human
breast~\cite{Arridge_1999}. The relatively low quality of
reconstructed images is due to severe ill-posedness of the inverse
problem.

Despite  significant recent progress in optical imaging of both
optically thin and thick media, little has been done for imaging of
systems of intermediate optical thickness. This represents the
subject of this study.  In radiative transport theory, such systems
are referred to as {\em mesoscopic}, meaning that the photon mean free
path (also known as the scattering length) is of order of the system
size~\cite{vanrossum_1999}. In the mesoscopic scattering regime,
applications to biological systems include engineered tissues and
semitransparent organisms such as zebra fish, animal embryos, or
small-animal extremities. In this case, light exhibits sufficiently
strong scattering so that the image reconstruction methods of computed
tomography are not applicable, yet the detected light is not diffuse
and the diffusion tomography can not be employed.

 In mesoscopic systems, the DA does not hold and the RTE must be used
 to describe the propagation of light~\cite{vanrossum_1999}. In this
 study, light transport in the mesoscopic regime is described by the
 first-order scattering approximation to the radiative transport
 equation.  This enables the derivation of a relationship between the
 extinction coefficient of the medium and the single-scattered light
 intensity, which represents the basis for a novel three-dimensional
 optical imagining technique that we propose and refer to as {\em
 single scattering optical tomography} (SSOT).  SSOT uses
 angularly-selective measurements of scattered light intensity to
 reconstruct the optical properties of macroscopically inhomogeneous
 media, assuming that the measured light is predominantly
 single-scattered. The image reconstruction problem of SSOT consists
 of inverting a generalization of the Radon transform in which the
 integral of the extinction coefficient along a broken ray (which
 corresponds to the path of a single-scattered photon) is related to
 the measured intensity.

Our results are
remarkable is several regards. First, similar to the case of computed
tomography, inversion of the broken-ray Radon transform is only mildly
ill-posed. Second, the inverse problem of the SSOT is two-dimensional
and three-dimensional image reconstruction can be performed
slice-by-slice. Third, in contrast to computed tomography, the
experimental implementation of SSOT does not require rotating the
imaging device around the sample to acquire data from multiple
projections. Therefore, SSOT can be used in the backscattering
geometry. Finally, SSOT makes use of intensity measurements, as
distinct from the more technically challenging experiments of optical
coherence microscopy or ISAM, which require information about the
optical phase. 


This paper is organized as follows. In Sec.~\ref{sec:math}, we
introduce the single-scattering approximation appropriate for the
mesoscopic regime of radiative transport. We then derive a
relationship between the scattering and absorption coefficients and
the single-scattered intensity. This relationship is then exploited in
Sec.~\ref{sec:principle} to discuss the physical principles of
SSOT. In Sec.~\ref{sec:simulations}, numerical algorithms for both the
forward and inverse problems are presented and illustrated in computer
simulations.

\section{Mesoscopic Radiative Transport}
\label{sec:math}

We begin by considering the propagation of light in a random medium of
volume $V$.  The specific intensity $I(\Br,\s)$ is the intensity
measured at the point $\bf r$ and in the direction $\s$, and is
assumed to obey the time-independent RTE
\begin{equation}
\label{RTE}
\left[ \hat{\bf s} \cdot \nabla + \mu_a({\bf r}) + \mu_s({\bf r}) \right] 
I({\bf r}, \hat{\bf s}) = 
\mu_s({\bf r}) \int A(\hat{\bf s},\hat{\bf s}^{\prime}) 
I({\bf r}, \hat{\bf s}^{\prime})d^2{s}^{\prime} \ , \ \ {\bf r}\in
V \ ,
\end{equation}
where $\mu_a({\bf r})$ and $\mu_s({\bf r})$ are the absorption and
scattering coefficients. The phase function $A(\hat{\bf s},\hat{\bf
s}^{\prime})$ describes the conditional probability that a photon
traveling in the direction $\hat{\bf s}$ is scattered into the
direction $\hat{\bf s'}$ and is normalized so that $\int A(\hat{\bf
s},\hat{\bf s}^{\prime}) d^2\hat{\bf s}^{\prime} = 1$ for all
$\s$. Eq.~(\ref{RTE}) is supplemented by a boundary condition of the
form

\begin{equation}
\label{RTE_bc_ihg}
I({\bf r},\hat{\bf s}) 
=  I_{\rm inc}({\bf r},\hat{\bf s}) \ , \ \ \hat{\bf s} \cdot \hat{\bf n}({\bf r}) < 0 \ , \ \ \Br \in \partial V \ ,
\end{equation}

\noindent
where $\hat{\bf n}$ is the outward unit normal to $\partial V$ and
$I_{\rm inc}$ is the incident specific intensity at the boundary.

The RTE (\ref{RTE}) together with the boundary
condition (\ref{RTE_bc_ihg}) can be equivalently formulated as the
integral equation

\begin{equation}
\label{RTE_int}
I({\bf r}, \hat{\bf s}) = I_b({\bf r}, \hat{\bf s}) + \int G({\bf r},\hat{\bf s}; {\bf r}^{\prime},\hat{\bf s}^{\prime}) \mu_s({\bf r}^{\prime}) A(\hat{\bf s}^{\prime},\hat{\bf s}^{\prime\prime})
I({\bf r}^{\prime}, \hat{\bf s}^{\prime\prime}) d^3r^{\prime} d^2
{s}^{\prime}d^2{s}^{\prime\prime} \ .
\end{equation}

\noindent
Here $I_b$ is the ballistic (unscattered) contribution to the specific
intensity and $G$ is Green's function for the ballistic RTE, which
satisfies the equation

\begin{equation}
\label{RTE_ballistic}
\left[ \hat{\bf s} \cdot \nabla + \mu_a({\bf r}) + \mu_s({\bf r}) \right] 
I_b({\bf r}, \hat{\bf s}) = 0 \ , 
\end{equation}

\noindent
and obeys the boundary condition (\ref{RTE_bc_ihg}). If a narrow
collimated beam of intensity $I_0$ is incident on the medium at the
point ${\bf r}_1$ in the direction $\hat{\bf s}_1$, then $I_b({\bf r},
\hat{\bf s})$ is given by

\begin{equation}
\label{I_b}
I_b({\bf r}, \hat{\bf s}) = I_0 G({\bf r},\hat{\bf s}; {\bf
  r}_1,\hat{\bf s}_1) \ ,
\end{equation}

\noindent
where ballistic Green's function $G({\bf r},\hat{\bf s}; {\bf
  r}^{\prime},\hat{\bf s}^{\prime})$ is expressed as

\begin{equation}
\label{G}
G({\bf r},\hat{\bf s}; {\bf r}^{\prime},\hat{\bf s}^{\prime}) = 
g({\bf r}, {\bf r}^{\prime})
\delta\left(\hat{\bf
    s}^{\prime} - \frac{\Br-\Br'}{|\Br-\Br'|}\right) \delta(\hat{\bf s} - \hat{\bf s}^{\prime}) . 
\end{equation}
Here
\begin{equation}
\label{g}
g({\bf r}, {\bf r}^{\prime}) = 
{1 \over {\vert {\bf r} - {\bf r}^{\prime}
\vert^2}} \exp\left[ - \int_0^{\vert {\bf r} - {\bf r}^{\prime}\vert}
\mu_t\left({\bf r}^{\prime} + \ell \frac{\Br-\Br'}{|\Br-\Br'|}\right)d\ell \right] \ ,
\end{equation}
and the Dirac delta function $\delta (\hat{\bf s}-\hat{\bf s}^{'})$ is
defined by
\begin{equation}
\delta (\hat{\bf s}-\hat{\bf s}^{'}) = \delta(\varphi_{\hat{\bf s}} -
\varphi_{\hat{\bf s}^{'}}) \delta\left( \cos\theta_{\hat{\bf s}} -
  \cos\theta_{\hat{\bf s}^{'}} \right ) ,
\end{equation}
where we have introduced the extinction (attenuation) coefficient
$\mu_t = \mu_a + \mu_s$, and $\theta$ and $\varphi$ are the polar
angles of the respective unit vectors. Note that $g$ is the
angularly-averaged ballistic Green's function,

\begin{equation}
\label{g_G}
g({\bf r}, {\bf r}^{\prime}) = \int d^2 {s} d^2 {s}^{\prime}
G({\bf r},\hat{\bf s}; {\bf r}^{\prime},\hat{\bf s}^{\prime}) \ .
\end{equation}

\noindent

By iterating Eq.~(\ref{RTE_int}) starting from $I^{(0)}=I_b$,
corresponding to ballistic light propagation, we obtain
\begin{equation}
\label{series}
I({\bf r}, \hat{\bf s})= I^{(0)}({\bf r}, \hat{\bf s})+ I^{(1)}({\bf
r}, \hat{\bf s})+ I^{(2)}({\bf r}, \hat{\bf s})+\cdots \ ,
\end{equation}
where  each term of the series is given by
\begin{equation}
I^{(n)}({\bf r}, \hat{\bf s})= \int d^3r^{\prime} d^2
{s}^{\prime}d^2{s}^{\prime\prime} G({\bf r},\hat{\bf s}; {\bf
  r}^{\prime},\hat{\bf s}^{\prime}) \mu_s({\bf r}^{\prime}) A(\hat{\bf
  s}^{\prime},\hat{\bf s}^{\prime\prime}) I^{(n-1)}({\bf r}^{\prime},
\hat{\bf s}^{\prime\prime}).
\end{equation}
The Born series (\ref{series}) can be regarded as an expansion in the
number of scattering events, each term corresponding to a successively
higher order of scattering \cite{duderstadt_book}. The convergence of
this series requires that $||\int d^3r^{\prime} d^2 {s}^{\prime}
G({\bf r},\hat{\bf s}; {\bf r}^{\prime},\hat{\bf s}^{\prime})
\mu_s({\bf r}^{\prime}) A(\hat{\bf s}^{\prime},\hat{\bf
s}^{\prime\prime})||_{\infty}<1$, where $||\cdot ||_{\infty}$ is the
$L^{\infty}$ norm. For an isotropically scattering random medium, this
norm can be calculated to be of the order \cite{duderstadt_book}
$(\mu_s/\mu_t)(1-\mbox{exp}(-\mu_t R))$, where $R$ is the
characteristic size of the system. Thus, the convergence requirement
for the Born series associated with the RTE is always satisfied. For
the system under investigation, this norm varies from $0.16$ to
$0.5$, as the amount of scattering is increased such that $\mu_s R$
varies from $1.6$ to $6.4$, respectively. Therefore, very rapid
convergence is expected.
\begin{figure}
\centerline{\input{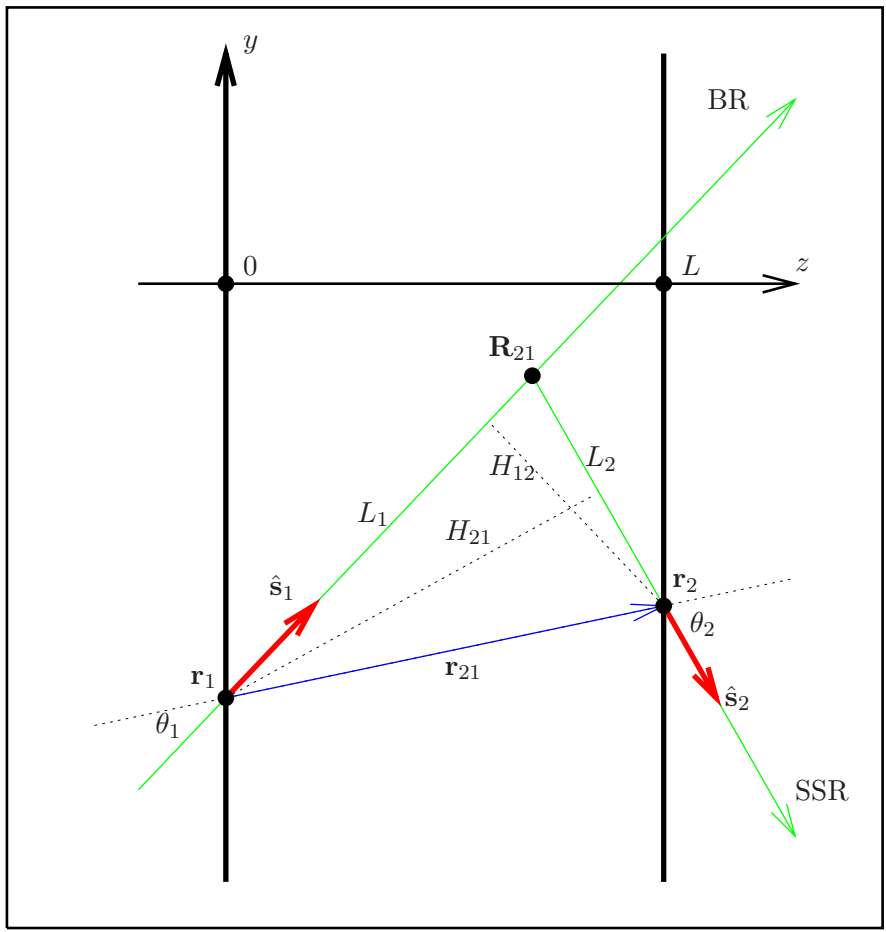}}
\caption{\label{fig:sketch2} (Color online) Illustrating the geometrical quantities used in Eq.~(\ref{I_s_2_1}). BR denotes broken ray.}
\end{figure}

In general, the specific intensity can be decomposed as $I=I_b+I_s$,
where $I_s$ is the scattered part of the specific intensity. Within
the accuracy of the single-scattering approximation, $I_s$ is given by
the expression
\begin{equation}
\label{I_s}
I_s({\bf r}, \hat{\bf s}) = \int d^3r^{\prime} d^2
{s}^{\prime}d^2{s}^{\prime\prime}G({\bf
  r},\hat{\bf s}; {\bf r}^{\prime},\hat{\bf s}^{\prime}) \mu_s({\bf
  r}^{\prime}) A(\hat{\bf s}^{\prime},\hat{\bf s}^{\prime\prime})
I_b({\bf r}^{\prime}, \hat{\bf s}^{\prime\prime})  \ .
\end{equation}
Note that the single scattering approximation is expected to hold in
the mesoscopic regime of radiative transport, where the system size is
of order the scattering length (defined to be $\mu_s^{-1}$).

We now assume that the sample is a slab of width $L$ and that the beam
is incident on one face of the slab at the point ${\bf r}_1$ in the
direction $\hat{\bf s}_1$ and that the transmitted intensity is
measured on the opposite face of the slab at the point ${\bf r}_2$ in
the direction $\hat{\bf s}_2$ (see Fig.~\ref{fig:sketch2}).  We denote
the scattered intensity measured in an such experiment by $I_s({\bf
r}_1, \hat{\bf s}_1;{\bf r}_2, \hat{\bf s}_2)$.  Performing the
integral in expression (\ref{I_s}) with $I_b$ given by (\ref{I_b}) and
using (\ref{G}) yields

\begin{eqnarray}
I_s({\bf r}_1, \hat{\bf s}_1;{\bf r}_2, \hat{\bf s}_2) = I_0
\Theta(\pi - \theta_1 - \theta_2) 
\delta\left( \vert \varphi_{\hat{\bf s}_1} - \varphi_{\hat{\bf s}_2}
  \vert - \pi \right) 
\frac{\mu_s({\bf R}_{21})A(\hat{\bf
        s}_2,\hat{\bf s}_1)}{r_{21} \sin \theta_1 \sin \theta_2}
    \nonumber \\ 
\times \exp\left[ - \int_0^{L_1}\mu_t({\bf r}_1 + \ell\hat{\bf s}_1)d\ell - \int_0^{L_2}\mu_t({\bf R}_{21} + \ell\hat{\bf s}_2)d\ell \right] \ , 
\label{I_s_2_1}
\end{eqnarray}

\noindent
where $\Theta(x)$ is the step function, ${\bf R}_{21}$ is the position
of the turning point of the ray, ${\bf r}_{21}={\bf r}_2 - {\bf
r}_1$, $r_{21}=\vert {\bf r}_{21} \vert$, $L_1 = \vert {\bf R}_{21} -
{\bf r}_1 \vert$, $L_2 = \vert {\bf r}_2 - {\bf R}_{21} \vert$, and
the angles $\theta_1$ and $\theta_2$ are defined by
$\cos\theta_{1,2}=\hat{\bf r}_{21}\cdot\hat{\bf s}_{1,2}$. The details of the derivation of Eq. (\ref{I_s_2_1}) are presented in the Appendix. We note the
following important relations:

\begin{equation}
\label{R_21_def}
{\bf R}_{21} = {\bf r}_1 + L_1 \hat{\bf s}_1 = {\bf r}_2 - L_2
\hat{\bf s}_2
\end{equation}

\begin{equation}
\label{L_1_2_def}
L_1 = r_{21} \frac{\sin\theta_2}{\sin(\theta_1 + \theta_2)} \ , \ \ 
L_2 = r_{21} \frac{\sin\theta_1}{\sin(\theta_1 + \theta_2)} \ ,
\end{equation}

The physical meaning of the various terms in (\ref{I_s_2_1}) is
as follows. First, the angles $\varphi_{\hat{\bf s}_{1,2}}$ in the
Dirac delta function $\delta\left( \vert \varphi_{\hat{\bf s}_1} -
  \varphi_{\hat{\bf s}_2} \vert - \pi \right)$ are the azimuthal
angles of the unit vectors $\hat{\bf s}_{1,2}$ in a reference frame whose
$z$-axis intersects both the position of the source and the detector
(this axis is shown by a dashed line in Fig.~\ref{fig:sketch2} and
should not be confused with the $z$-axis of the laboratory frame shown
by a solid line). The presence of this
one-dimensional delta function is the manifestation of the fact that
two straight rays exiting from the points ${\bf r}_1$ and ${\bf r}_2$
in the directions $\hat{\bf s}_1$ and $-\hat{\bf s}_2$, respectively,
can intersect only if $\hat{\bf s}_1$ and $\hat{\bf s}_2$ and ${\bf
r}_{21}$ are in the same plane (equivalently, if $\varphi_{\hat{\bf
s}_1} - \varphi_{\hat{\bf s}_2}=0, \pm \pi$) and point into
different half-planes (this requires that $\varphi_{\hat{\bf s}_1} -
\varphi_{\hat{\bf s}_2}=\pm \pi$). Second, the point of intersection
exists within the plane only if $\theta_1+\theta_2 < \pi$, which is
expressed by the step function $\Theta(\pi - \theta_1 - \theta_2)$. We
note that if, additionally, $\hat{\bf s}_1$ and $\hat{\bf s}_2$ are
restricted so that $\hat{\bf z} \cdot \hat{\bf s}_1>0$ and $\hat{\bf
  z} \cdot \hat{\bf s}_2<0$ ($\hat{\bf s}_1$ points into the slab and
$\hat{\bf s}_2$ points out of the slab), then ${\bf R}_{21}$ lies within
the slab.  Third, the factor $\mu_s({\bf R}_{21})A(\hat{\bf s}_2,
\hat{\bf s}_1)$ is the ``probability'' that the ray is scattered at
the point ${\bf r}={\bf R}_{21}$ and changes direction from $\hat{\bf
s}_1$ to $\hat{\bf s}_2$. This factor is, in general,
position-dependent.  Fourth, $1/r_{21}\sin\theta_1\sin\theta_2$ is a
geometrical factor.  We note that it can be equivalently rewritten as
$r_{21}/H_{21}H_{12}$, where $H_{21}$ and $H_{12}$ are the two heights
of the triangle $({\bf r}_1,{\bf R}_{21},{\bf r}_2)$ drawn from the
vertices ${\bf r}_1$ and ${\bf r}_2$, respectively, as shown in
Fig.~\ref{fig:sketch2}. Finally, it can be seen that the integral of $\mu_t$ in
the argument of the exponential is evaluated along a broken ray which begins
at $\Br_1$, travels in the direction $\s_1$, turns at the point $\BR_{21}$, travels in the direction $\s_2$, and exits the slab at $\Br_2$.

\section{Physical principles of SSOT}
\label{sec:principle}

\begin{figure*}
\includegraphics*[ width=0.45\linewidth]{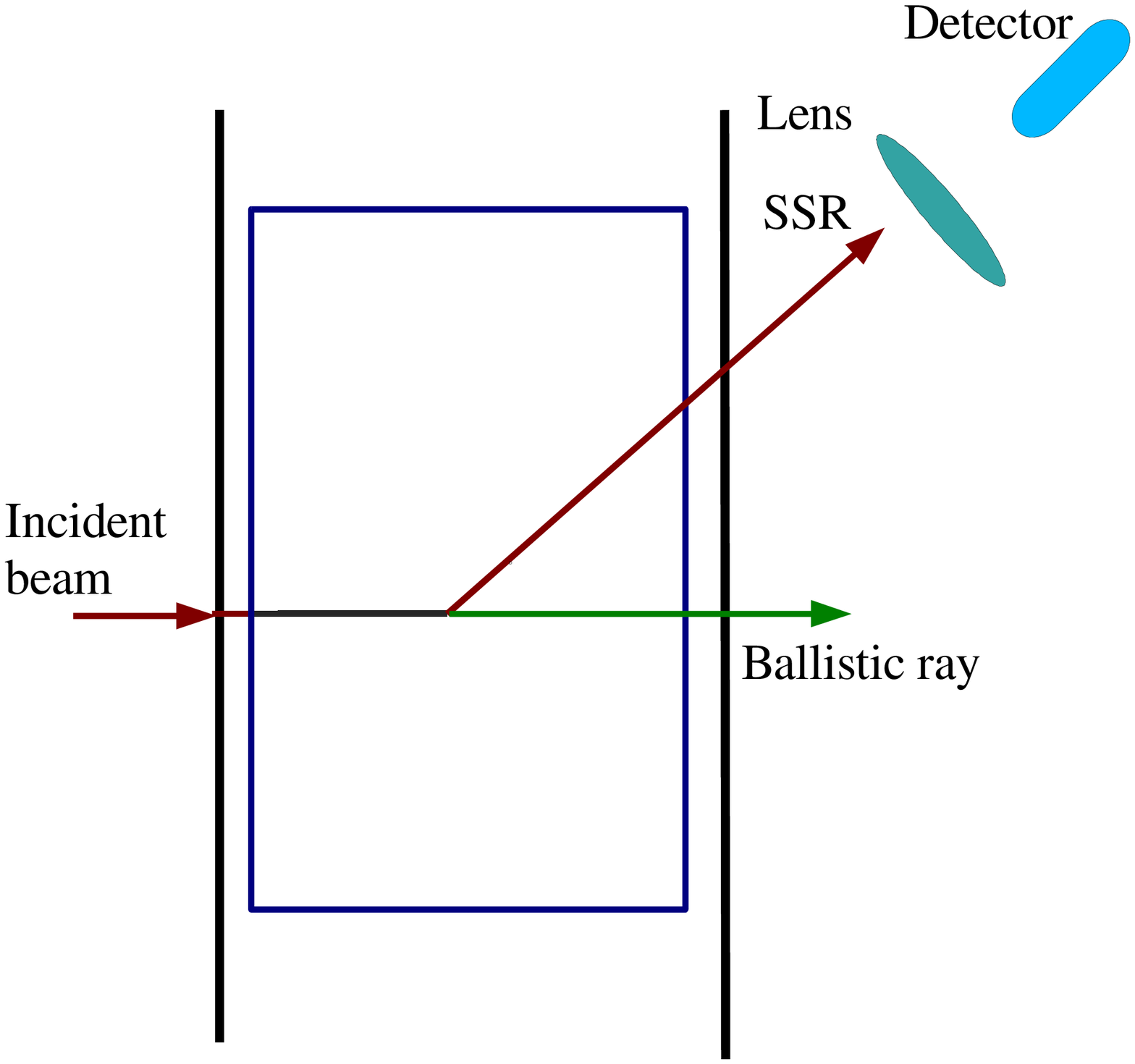}
\includegraphics*[ width=0.45\linewidth]{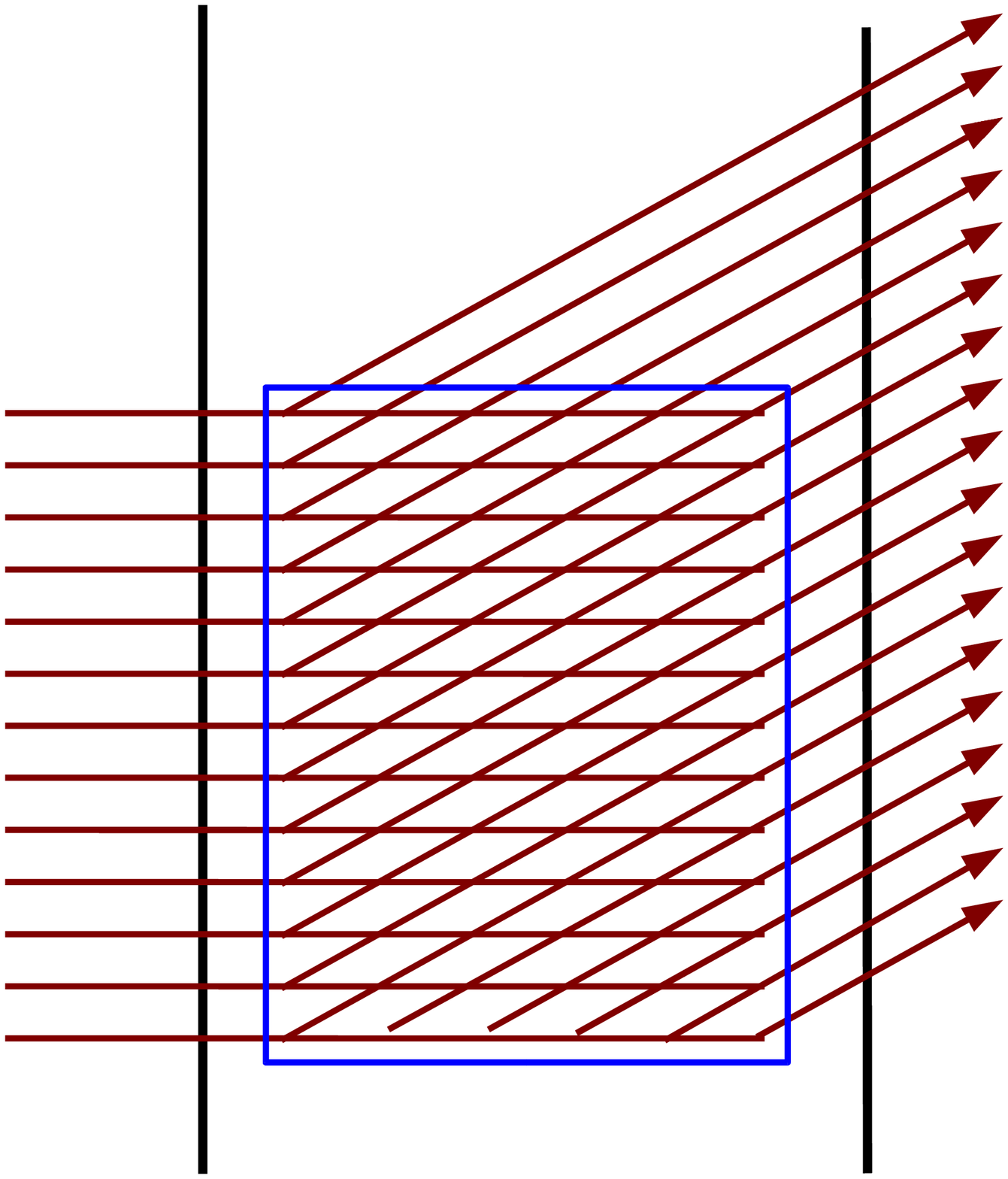}
\caption{(Color online)  (a) Schematic illustration of
the broken-ray (or single-scattered ray, denoted here by SSR) trajectory. 
(b) Source-detector
arrangement for SSOT. Reconstruction is performed independently
slice-by slice. The blue rectangle represents the area in which a
reconstruction can be performed.}
\label{fig:sketch1}
\end{figure*}

The physical principle of SSOT is illustrated in
Fig.~\ref{fig:sketch1} where a slab-shaped sample is illuminated by a
normally incident beam. In the absence of scattering, the beam would
propagate ballistically, as shown by the green ray.  Detection of such
unscattered rays is the basis of computed tomography. In the presence
of scattering, the ray can change direction as shown in
Fig.~\ref{fig:sketch1}(a).  Of course, scattering does not result in
elimination of the ballistic ray.  However, it is possible to employ
angularly-selective detectors that do not register the ballistic
component of the transmitted light. In the example shown in
Fig.~\ref{fig:sketch1}(a), it is assumed that only the intensity of
the broken ray shown by the red line is detected.  To avoid detection
of the ballistic component of the transmitted light, the
angularly-selective source and detector are not aligned with each
other. Moreover, the data can be collected either on opposite sides of
the slab (transmission measurements), or in the backscattering
geometry.  In both cases, rotation of the instrument around the sample
is not required.

By utilizing multiple incident beams and detecting the light exiting
the sample at different points, as shown in Fig.~\ref{fig:sketch1}(b),
we will see that it is possible to collect sufficient data to
reconstruct the spatial distribution of the attenuation coefficient in
a fixed transverse slice of the slab.  In addition to varying the
source and detector positions, one can also vary the angles of
incidence and detection. In principle, this can provide additional
information for simultaneous reconstruction of absorption and
scattering coefficients of the sample, a topic we will consider
elsewhere.

The image reconstruction problem of SSOT is to reconstruct $\mu_a$,
$\mu_s$ and $A$ from measurements of $I_s$. For simplicity, we assume
that $\mu_s$ and $A$ are known, in which case we wish to determine
$\mu_a$. To proceed, we use  Eq.~(\ref{I_s_2_1}) to separate the known or measured quantities from those that need to be reconstructed, and  introduce the data function

\begin{equation}
\label{phi_def}
\phi({\bf r}_1,\hat{\bf s}_1;{\bf r}_2,\hat{\bf s}_2) = 
-\ln\left[ \frac{r_{21} \sin\theta_1 \sin\theta_2 \int 
I_s({\bf r}_1,\hat{\bf s}_1;{\bf r}_2,\hat{\bf s}_2)
d\varphi_{\hat{\bf s}_2}}{I_0 \mu_s  A(\hat{\bf s}_1,\hat{\bf s}_2)}
\right] \ ,
\end{equation}
where we have assumed that $A$ is position-independent.   Note that if
$I_s$ is experimentally measured, the angular integration on the
right-hand side of (\ref{phi_def}) does not need to be performed
numerically. The measured data is necessarily integrated in a narrow
interval of $\varphi_{\hat{\bf s}_2}$ due to the finite aperture and
acceptance angle of the detector. Note also that the above definition
is only applicable for such configurations of sources and detectors
such that $\theta_1+\theta_2<\pi$. Otherwise, any measured intensity
is due to higher-order terms in the collision expansion which are not
accounted for (\ref{I_s_2_1}).

Making use of the definition (\ref{phi_def}) and Eq.~(\ref{I_s_2_1}), we find that
$\mu_t$ obeys the integral equation

\begin{equation}
\label{Ray_int_3}
\phi({\bf r}_1,\hat{\bf s}_1;{\bf r}_2,\hat{\bf s}_2) = \int_{{\rm BR}({\bf r}_1,\hat{\bf s}_1;{\bf r}_2,\hat{\bf s}_2)}\mu_t({\bf r}(\ell)) d\ell \ .
\end{equation}
Here the integral is evaluated along the broken ray
(such as the one shown in Fig.~\ref{fig:sketch2}), which is uniquely defined by
the source and detector positions and orientations, and $\ell$ is the
linear coordinate along the ray.

According to (\ref{Ray_int_3}), the attenuation function is linearly
related to the data function. In this respect, the mathematical
structure of SSOT is similar to the problem of inverting the Radon
transform in computed tomography except that the integrals are
evaluated along broken rays.  Since $\mu_t$ may be regarded as a
function of two variables, it is sufficient to consider only
two-dimensional measurements. One possible choice is to vary the
source and detector coordinates, $y_1$ and $y_2$, while keeping the
angles of incidence and detection $\beta_1$ and $\beta_2$ fixed. By
$\beta_1$ and $\beta_2$ we mean here the angles between the $z$-axis
of the laboratory frame and the unit vectors $\hat{\bf s}_1$ and
$\hat{\bf s}_2$, respectively.  Note that these angles are not equal
to the angles $\theta_1$ and $\theta_2$ shown in
Fig.~\ref{fig:sketch2}. The latter can vary in the measurement scheme
described in this section, while $\beta_1$ and $\beta_2$ are fixed.
Below, we omit $\beta_1$ and $\beta_2$ from the list of formal
arguments of the data function and consider the equation

\begin{equation}
\label{SSOT_2D}
\phi(y_1,y_2)= \int_{{\rm BR}(y_1,y_2)} \mu_t(y(\ell), z(\ell)) d\ell \ ,
\end{equation}

\noindent
where $\phi(y_1,y_2)$ is the two-dimensional data function.

As explained above, the selection of the points and directions of
incidence and detection define a slice in which $\mu_t$ is to be
reconstructed. In Fig.~\ref{fig:sketch2}, this slice coincides with
the $yz$-plane of the laboratory frame. Assuming that the
$x$-coordinate is fixed, we then regard $\mu_t$ as a function of $y$
and $z$. Three-dimensional reconstruction is then performed
slice-by-slice.

\section{Image Reconstruction}
\label{sec:simulations}

In this section we illustrate image reconstruction in SSOT using a
numerical technique based on discretization and algebraic inversion of
the two-dimensional integral equation (\ref{SSOT_2D}). We note that
more sophisticated image reconstruction procedures which utilize the
translational invariance of rays may also be derived. These methods
are conceptually similar to those previously developed for optical
diffusion tomography~\cite{schotland_01_1,markel_04_4} and will be
described elsewhere.

\subsection{Forward Problem}
\label{subsec:forward}

We begin by describing a method to generate simulated forward data to
test the SSOT image reconstruction. Assuming an isotropically
scattering sample with $A(\hat{\bf s}, \hat{\bf s}') = 1/4\pi$, it can
be shown from (\ref{RTE_int})~\cite{erdmann_68_1,larsen_74_1} that the
specific intensity everywhere inside the sample is related to the
density of electromagnetic energy $u({\bf r})\equiv \int I({\bf
  r},\hat{\bf s}) d^2{s}$ by the formula

\begin{equation}
\label{I_u}
I({\bf r},\hat{\bf s}) = I_b({\bf r},\hat{\bf s}) + \frac{1}{4\pi}\int G({\bf
  r},\hat{\bf s}; {\bf r}^{\prime},\hat{\bf s}^{\prime})
\mu_s({\bf r}^{\prime})u({\bf r}^{\prime}) d^3r^{\prime}
d^2{s}^{\prime},
\end{equation}

\noindent
where $u({\bf r})$ satisfies the integral equation

\begin{equation}
\label{u_u_b}
u({\bf r}) = u_b({\bf r}) + \frac{1}{4\pi}\int g({\bf r},{\bf r}^{\prime})
\mu_s({\bf r}^{\prime})u({\bf r}^{\prime}) d^3r^{\prime}
\ , 
\end{equation}

\noindent
Here $g({\bf r}, {\bf r}^{\prime})$ is given by (\ref{g}) and
$u_b({\bf r})\equiv \int I_b({\bf r},\hat{\bf s})d^2s$ is the
ballistic  energy density. We note that the assumption of isotropic
scattering is the most stringent test for SSOT.

The specific intensity is computed by first solving the integral
equation (\ref{u_u_b}) and then substituting the obtained solution
$u({\bf r})$ into (\ref{I_u}) (where the ballistic part $I_b$ may be
ignored). Note that $I({\bf r},\hat{\bf s})$ calculated from
(\ref{I_u}) satisfies the boundary conditions at all surfaces. We also
stress that this numerical approach is non-perturbative and includes
all orders of scattering.

In the simulations shown below, (\ref{u_u_b}) is discretized on a
rectangular grid and solved by standard methods of linear algebra. The
energy density $u({\bf r})$ is assumed constant within each cubic
cell, and the corresponding values $u_n=u({\bf r}_n)$, where ${\bf
r}_n$ is the center of the $n$th cubic cell, obey the algebraic system
of equations obtained by discretizing Eq.~(\ref{u_u_b}). The
off-diagonal elements of the matrix of this system corresponding to
the integral on the right-hand side of (\ref{u_u_b}) are given by
$(\mu_s h^3 / 4\pi) g({\bf r}_m,{\bf r}_n)$, where $h$ is the
discretization step. Here we can take advantage of the fact that
$\mu_s$ was set to be constant throughout the sample, while the
inhomogeneities were assumed to be purely absorbing.  Computation of
the diagonal elements is slightly more involved because $g({\bf
r},{\bf r}^{\prime})$ diverges when ${\bf r}\rightarrow {\bf
r}^{\prime}$.  In this case, we need to find an approximation for the
integral

\begin{equation}
\label{diag_int}
S = \frac{\mu_s}{4\pi} \int_{V_n} g({\bf r}_n,{\bf r})d^3r \ ,
\end{equation}

\noindent
where the integration is carried out over the $n$th cell. While
integration over a cubic volume is difficult, the important fact is
that the singularity in $g({\bf r},{\bf r}^{\prime})$ is integrable.
We then write, approximately,

\begin{equation}
\label{diag_int_approx}
S \approx \mu_s \int_0^{R_{\rm eq}} g(0,{\bf r})r^2 dr \ ,
\end{equation}

\noindent
where $R_{\rm eq}=(3/4\pi)^{1/3}h$ is the radius of a sphere of
equivalent volume. For a sufficiently fine disctretization, $\mu_t
R_{\rm eq} \ll 1$, which allows us to write $g(0,{\bf r})\approx
1/r^2$. This leads to $S = \mu_s R_{\rm eq}$, and the discretized
version of Eq.~(\ref{u_u_b}) becomes

\begin{equation}
\label{u_u_b_disc}
(1 - \mu_s R_{\rm eq}) u_n - \frac{\mu_s h^3}{4\pi}\sum_{m\neq n}g({\bf
  r}_n, {\bf r}_m) u_m = u_b({\bf r}_n) \ .
\end{equation}
\noindent
We note that this set of equations is an accurate approximation to the
integral equation (\ref{u_u_b}) if $\mu_s R_{\rm eq} \sim \mu_s h \ll
1$. However, since in practice this inequality may be not very strong,
the term $\mu_s R_{\rm eq}$ on the left-hand side of
(\ref{u_u_b_disc}) is not neglected.

Eq.~(\ref{u_u_b_disc}) can be written in matrix notation as
\begin{equation}
\label{u_u_b_matrix}
{\mathcal W} \vert u \rangle = \vert b \rangle
\end{equation}
where 
\begin{eqnarray}
\langle n \vert {\mathcal W} \vert m \rangle &=&
\frac{1}{\alpha}\delta_{nm} \ \nonumber \\ &&- \ \frac{h^2}{\vert {\bf
r}_n - {\bf r}_m \vert^2} \exp\left[ - \int_{0}^{\vert {\bf r}_n -
{\bf r}_m \vert} \mu_t \left( {\bf r}_m + \hat{\bf u}({\bf r}_n - {\bf
r}_m)\ell \right) d\ell \right] (1-\delta_{nm})\ ,
\label{W_def}
\end{eqnarray}
with $\delta_{nm}$ being the Kronecker delta function, $\alpha$ being  a
dimensionless coupling constant defined by

\begin{equation}
\label{alpha_def}
\alpha = \frac{\mu_s h}{4\pi (1 - \mu_s R_{\rm eq})} \ ,
\end{equation}

\noindent
$\langle n \vert u \rangle = u_n$, and  $\langle n \vert b \rangle =
4\pi u_b({\bf r}_n) / \mu_s h$. Note that the quantity $u_b({\bf
  r}_n)$ is defined as an average over the $n$th cell, namely,
$u_b({\bf r}_n) = h^{-3}\int_{V_n} u_b({\bf r}) d^3r$.

\begin{figure}
\centerline{\input{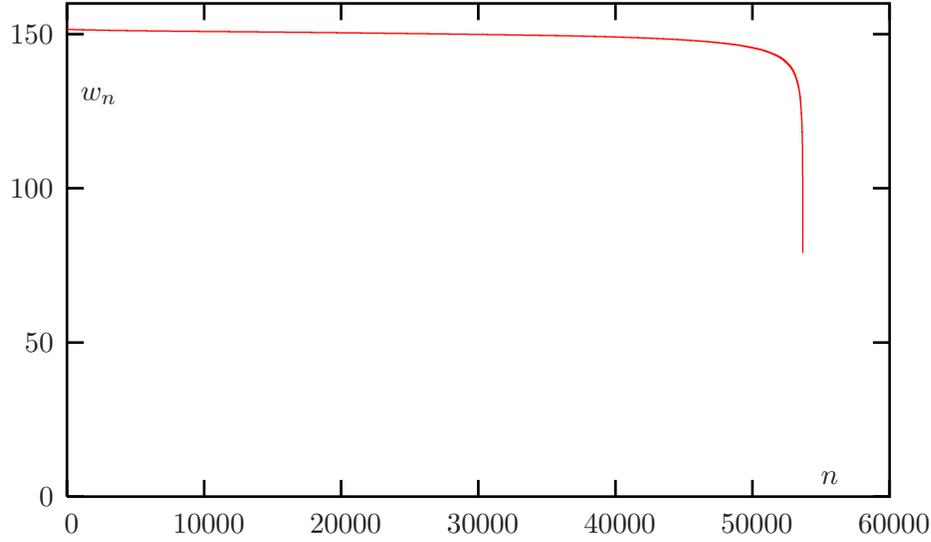}}
\caption{\label{fig:wf} (Color online) All $53,680$ eigenvalues $w_n$
  of the matrix ${\mathcal W}$ of Eq.~\ref{u_u_b_matrix} computed for
  $\mu_s=0.08h^{-1}$ and the set of absorbing inhomogeneities
  described in section~\ref{subsec:inv}.}
\end{figure}

Computing the elements of the matrix ${\mathcal W}$ requires the
evaluation of the integrals on the right-hand side of
(\ref{W_def}). For a homogeneous sample, this can be performed
analytically. For an inhomogeneous sample, it is done
numerically. Once this is accomplished, (\ref{u_u_b_matrix}) is then
solved by matrix inversion.  We note that the symmetric matrix
${\mathcal W}$ is well-conditioned \cite{condition-number}. This is
illustrated in Fig.~\ref{fig:wf} where we plot all the eigenvalues of
${\mathcal W}$ for $\mu_s=0.08h^{-1}$ with the set of absorbing
inhomogeneities described in section~\ref{results}.  We also note that
${\mathcal W}$ is positive-definite, so that $u$ and $u_b$ are
positive. Also, it was verified in all simulations that the diffuse
component of the density, defined as the quantity $u - u_b$, was
positive everywhere inside the sample.

Special attention must be given to the effects of discretization when
computing the single-scattered intensity, the data function, and the
forward data.  Since the computations involve discrete rays, employing
the expressions (\ref{I_s_2_1}) and (\ref{phi_def}) that
contain the delta-function $\delta\left( \vert \varphi_{\hat{\bf s}_1}
- \varphi_{\hat{\bf s}_2} \vert - \pi \right)$ and the geometrical
factor $1/r_{21} \sin\theta_2 \sin\theta_1$ becomes
cumbersome. Instead, we have  derived the discrete analogues of these
expressions starting form Eq.~(\ref{I_s}) from which (\ref{I_s_2_1})
has  been obtained.  In particular, the expression for the
single-scattered intensity (the discrete analog of (\ref{I_s_2_1})) is
obtained from Eq.~(\ref{I_s}) as
\begin{equation}
\label{I_s_2_1_d}
I_s({\bf r}_2, \hat{\bf s}_2;{\bf r}_1, \hat{\bf s}_1) = \frac{\mu_s
  h^3}{4\pi} g({\bf r}_2, {\bf R}_{21}) u_b({\bf R}_{21}) \ ,
\end{equation}
\noindent
where $u_b({\bf R}_{21})$ is the average of $u_b({\bf r})$ over the
cell that contains ${\bf R}_{21}$. This expression and the expression
(\ref{I_b}) for the ballistic intensity suggest the definition of the data function of the form
\begin{equation}
\label{data_disctrete}
\phi({\bf r}_2, \hat{\bf s}_2;{\bf r}_1, \hat{\bf
  s}_1)=-{\mbox ln}\left[\frac{4\pi}{h^3}\frac{I_s({\bf r}_2, \hat{\bf
      s}_2;{\bf r}_1, \hat{\bf s}_1)}{I_0\mu_s}\right].
\end{equation}
This is the discrete analogue of (\ref{phi_def}).  Finally, the data
function is calculated according with this expression and using the
specific intensity obtained from the discretized version of
(\ref{I_u}),
\begin{equation}
\label{I_u_d}
I({\bf r}_2, \hat{\bf s}_2;{\bf r}_1, \hat{\bf s}_1) = \frac{\mu_s h^3}{4\pi} 
\sum_{{\bf r}_2-{\bf r}_n=\hat{\bf s}_2|{\bf r}_2-{\bf r}_n|} g({\bf r}_2, {\bf r}_n) u_n \ ,
\end{equation}

\noindent
where $u_n$ must be computed numerically for the selected source.  The
condition on the sum means that summation is performed only over cells
that are intersected by the ray exiting from the detection point ${\bf
r}_2$ in the direction $\hat{\bf s}_2$. The above formula is valid for
the specific measurement scheme which obtains when the intersection
length of all such rays with any cubic cell is constant. Otherwise, a
more complicated numerical integration must be employed. We note while
Eqs.~(\ref{data_disctrete}) and (\ref{I_u_d}) are employed for
simulated data, Eqs.~(\ref{I_s_2_1}) and (\ref{phi_def}) must be
used when experimental data is available.
Also, the use of experimental data avoids many mathematical
complications that arise due to discretization of rays, needed to
solve the forward RTE problem, usually computationally intensive.

In order to model the noise in measured data, the specific intensity
obtained from the forward solver was scaled and rounded so that it
was represented by 16-bit unsigned integers, similar to the
measurement by a typical CCD camera. Then a statistically-independent
positively-defined random variable was added to each measurement. The
random variables were evenly distributed in the interval $[0,n I_{\rm
  av}]$, where $n$ is the noise level indicated in the figure legends
below and $I_{\rm av}$ is the average measured intensity (a 16-bit
integer). The DC part (the positive background) of the intensity
was not subtracted (this procedure is commonly applied to the digital
output of CCD chips).  Then the simulated intensity measurements,
together with the appropriately scaled incident intensity $I_0$ were
substituted into (\ref{phi_def}) to obtain the data function $\phi$.

\subsection{Inverse Problem}
\label{subsec:inv}   

We now describe the method by which we invert the integral equation
(\ref{SSOT_2D}). The discrete version of (\ref{SSOT_2D}) has the form

\begin{equation}
\label{SSOT_2D_d}
\sum_n {\mathcal L}_{\nu n} \mu_{tn} = \phi_\nu \ ,
\end{equation}

\noindent
where the same grid is employed as is used to solve the forward
problem, ${\mathcal L}_{\nu n}$ is the length of the intersection of
the broken ray indexed by $\nu=(y_1,y_2)$ with the $n$th cubic cell
(located within the selected $x$-slice of the sample). The matrix
elements ${\mathcal L}$ are determined from simple geometric
considerations.  The matrix form of (\ref{SSOT_2D_d}) is

\begin{equation}
\label{SSOT_2D_matrix}
{\mathcal L}\vert \mu_t\rangle = \vert \phi \rangle \ . 
\end{equation}

\noindent
The equation (\ref{SSOT_2D_matrix})  can be solved using a
regularized pseudoinverse~\cite{natterer_book_01}, namely,

\begin{equation}
\label{mu_SVD}
\vert \mu_t^+ \rangle = ({\mathcal L}^* {\mathcal L})^{-1} {\mathcal
  L}^* \vert \phi \rangle \ .
\end{equation}

\begin{figure}
\centerline{\input{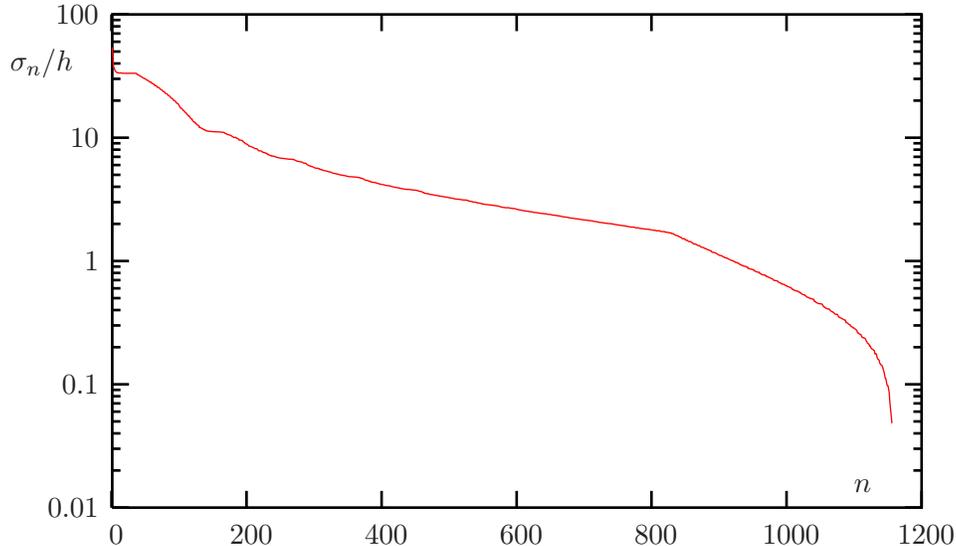}}
\caption{\label{fig:wi}(Color online) All $1,156$ non-zero singular values
  $\sigma_n$ of the matrix ${\mathcal L}$ defined by Eq.~(\ref{SSOT_2D_d}) (the size of ${\mathcal L}$
  in this example is $1,600 \times 1,156$).}
\end{figure}

\noindent
Here $({\mathcal L}^* {\mathcal L})^{-1}$ is understood in the
following sense:

\begin{equation}
\label{L_star_L_inv}
({\mathcal L}^* {\mathcal L})^{-1} = \sum_n \Theta(\sigma_n^2 -
\epsilon) \frac {\vert f_n \rangle \langle f_n \vert} {\sigma_n^2} \ ,
\end{equation}

\noindent
where $\Theta(x)$ is the step function, $\epsilon$ is a small
regularization parameter, and $\vert f_n \rangle$ and $\sigma_n$ are
the singular vectors and singular values ~\cite{natterer_book_01},
respectively, of the matrix ${\mathcal L}$. These quantities are the
solution of the symmetric eigenproblem ${\mathcal L}^* {\mathcal
  L}\vert f_n \rangle = \sigma_n^2 \vert f_n \rangle$. A typical
spectrum of singular values of ${\mathcal L}$ for $\mu_s=0.08h^{-1}$,
$1,600$ measurements and $34^2=1,156$ unknown values of $\mu_t$ (the
size of ${\mathcal L}$ in this example is $1,600 \times 1,156$, so
that the problem is slightly overdetermined) is shown in
Fig.~\ref{fig:wi}. It can be seen that matrix condition number
\cite{condition-number} for the inverse problem is much larger than
for the forward problem. In the example shown in Fig.~\ref{fig:wi},
the condition number is $\approx 10^3$. Thus the inverse problem is
very mildly ill-posed.

\subsection{Numerical Results}
\label{results}

In what follows, we illustrate applications of SSOT to biological
imaging. In particular, the numerical simulations presented here are
relevant for  ``semi-transparent'' systems, such as zebra
fish or engineered tissues. Also, the physical situations analyzed
here are experimentally encountered for organ tissues at certain
wavelengths of the illuminating beam
\cite{welch_book}.

\begin{figure}
\begin{center}
\includegraphics[height=19cm]{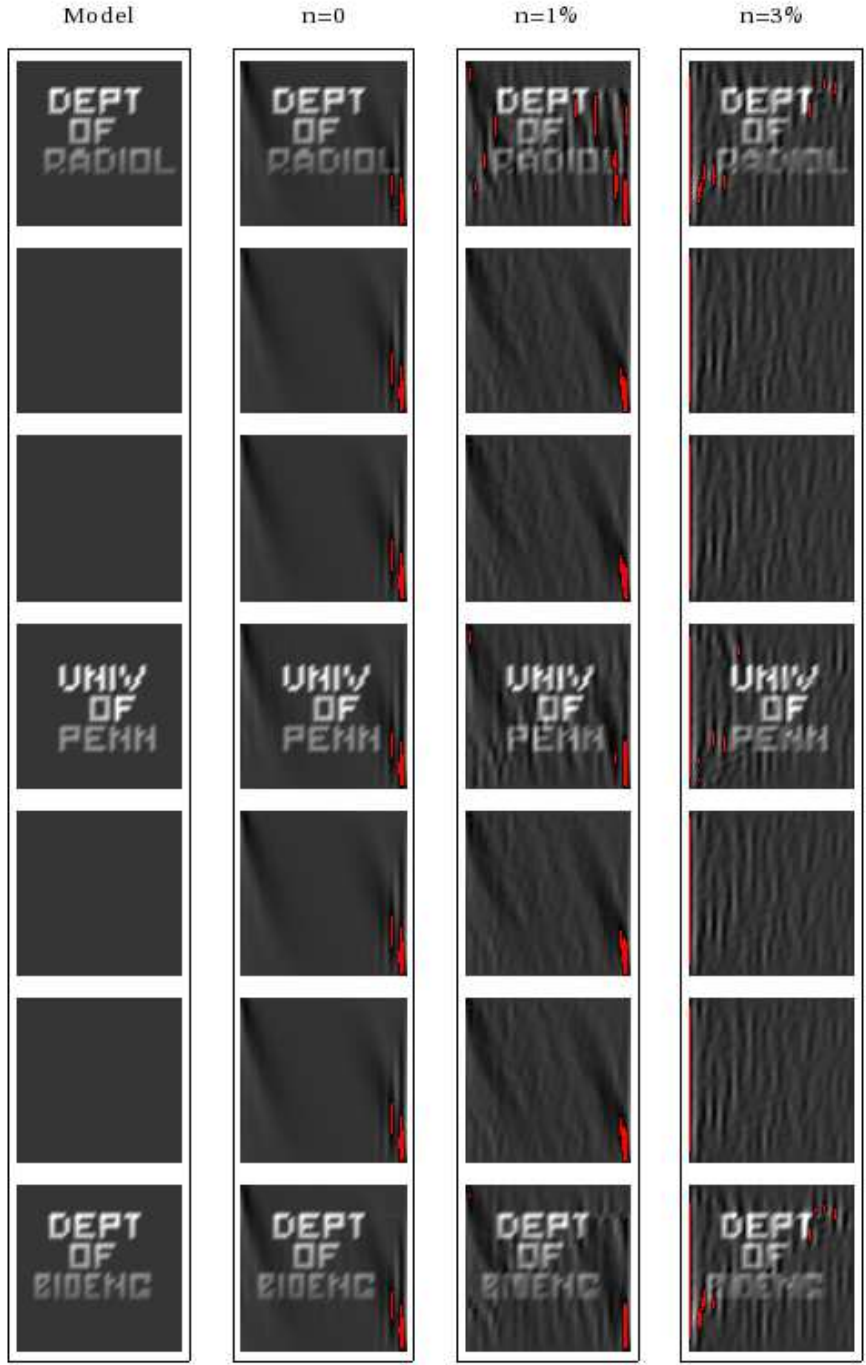}
\end{center}
\caption{\label{mu_s=0.04}(Color online) Image reconstruction for a slab with
$\mu_s=0.04h^{-1}$ and various noise levels $n$. The absorbing
inhomogeneities are placed in the slices $x=6h$, $x=13h$ and $x=20h$,
and the rows show the slices $x=6h$, $7h$, $12h$, $13h$, $14h$,
$19h$, and $20h$. The same color scale is used for all slices, with
the maximum (white) corresponding to $\mu_t=0.24h^{-1}$ and the
minimum (black) to $\mu_t=0$. Red  regions correspond to
negative values of the reconstructed extinction coefficient.}
\end{figure}

\begin{figure}
\begin{center}
\includegraphics[height=19cm]{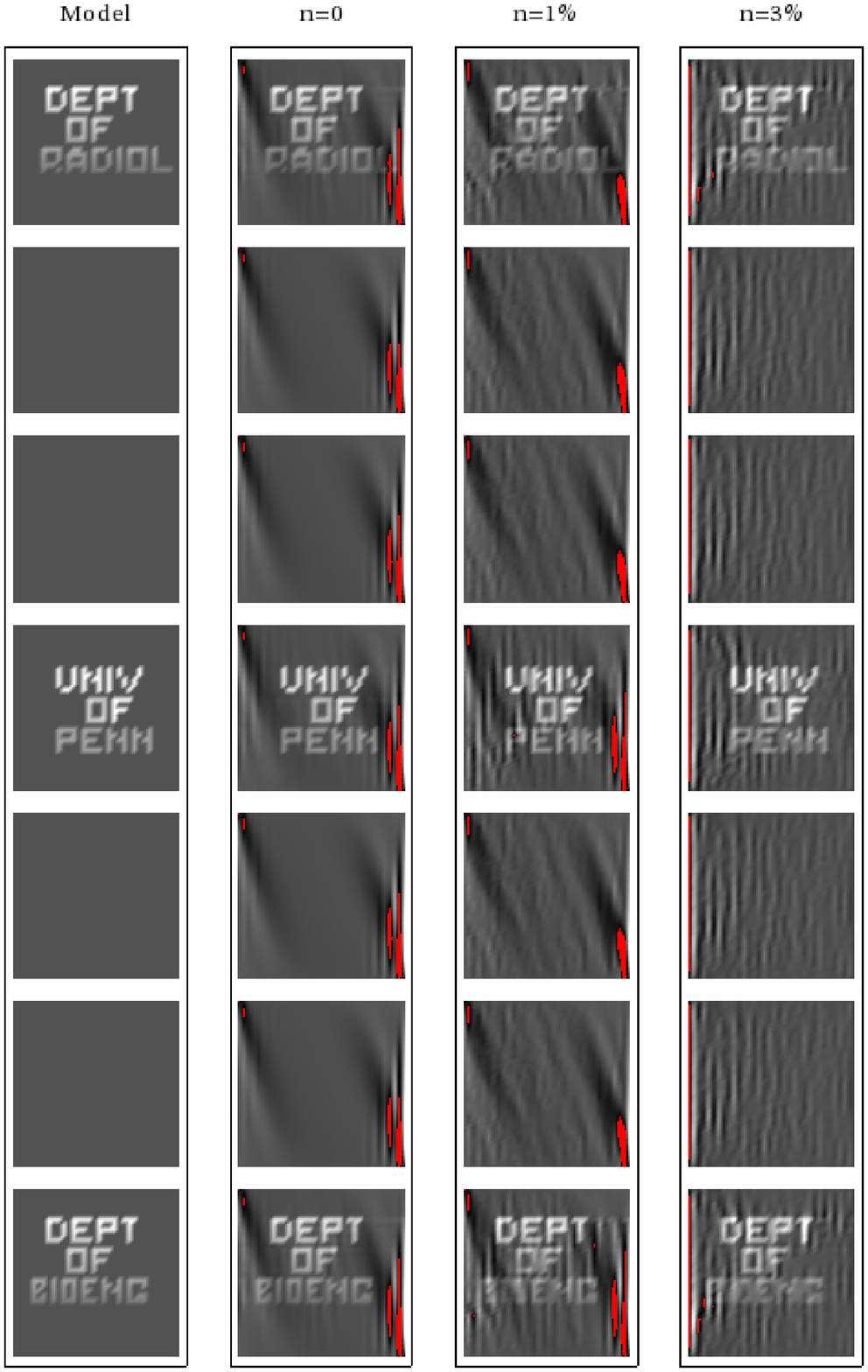}
\end{center}
\caption{\label{mu_s=0.08}(Color online) Image reconstruction for a
slab with $\mu_s=0.08h^{-1}$.  The same
color scale is used for all slices, with the maximum (white)
corresponding to $\mu_t=0.28h^{-1}$ and the minimum (black) to
$\mu_t=0$. All the other details are as for Fig.~\ref{mu_s=0.04}.}
\end{figure}

\begin{figure}
\begin{center}
\includegraphics[height=19cm]{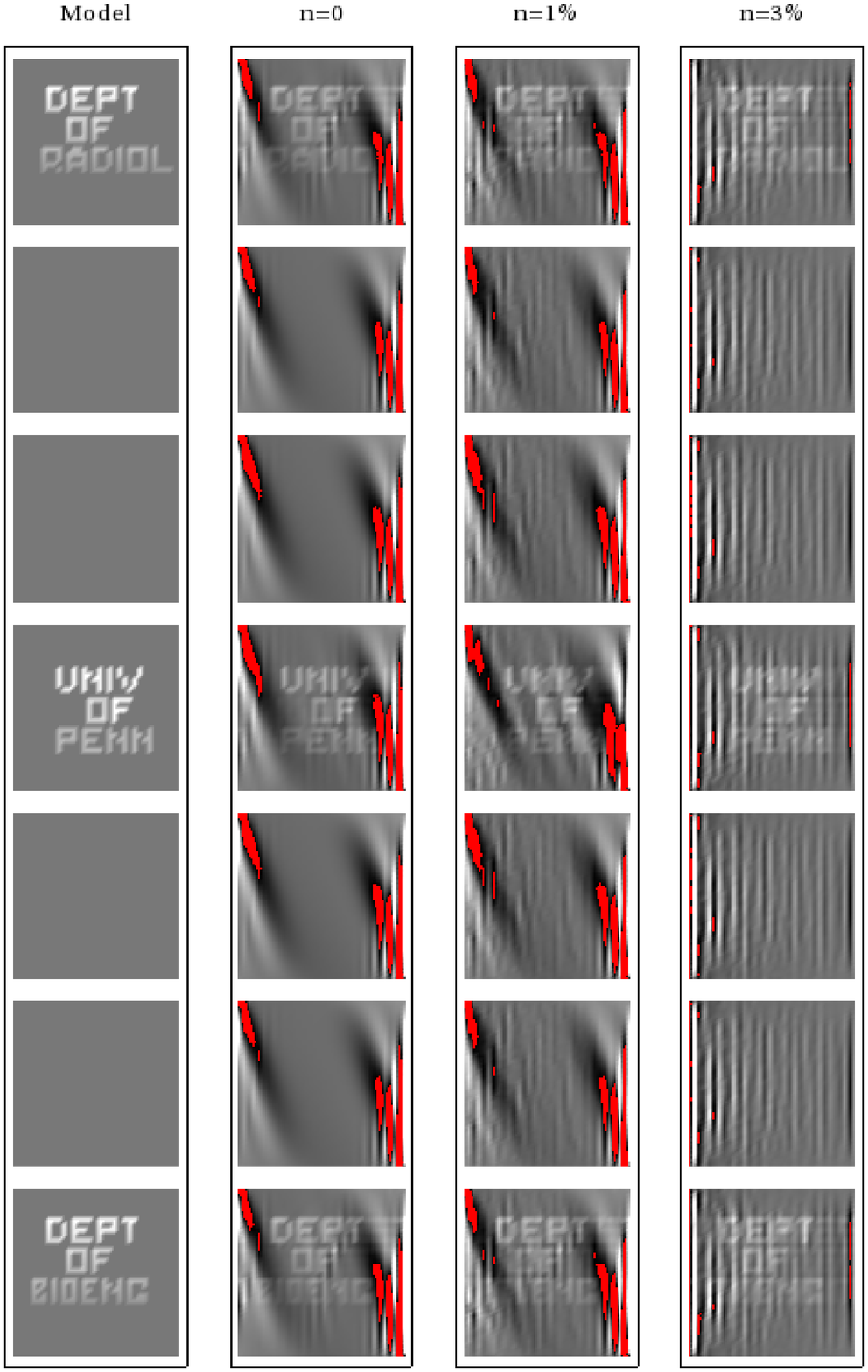}
\end{center}
\caption{\label{mu_s=0.16}(Color online) Image reconstruction for a
slab with $\mu_s=0.16h^{-1}$.  The same
color scale is used for all slices, with the maximum (white)
corresponding to $\mu_t=0.36h^{-1}$ and the minimum (black) to
$\mu_t=0$.  All the other details are as for Fig.~\ref{mu_s=0.04}.}
\end{figure}

 Reconstructions were carried out for a rectangular isotropically
scattering sample of dimensions $L_x=25h$, $L_y=122h$ and
$L_z=40h$. The background absorption coefficient of the sample was
equal to $0.01h^{-1}$ and was spatially modulated by absorbing
inhomogeneities (the target). The target was a set of absorbing
inclusions formed in the shape of letters, with absorption varying
from $0.06h^{-1}$ to $0.2h^{-1}$.  The inclusions were concentrated
only in three layers: $x=6h$, $x=13h$ and $x=20h$, as shown in the
columns marked model of Figs.~\ref{mu_s=0.04}-\ref{mu_s=0.16}.  The
scattering coefficient was constant throughout the sample, with three
different values used throughout the simulations corresponding to
$\mu_s=0.04h^{-1}$, $\mu_s=0.08h^{-1}$ and $\mu_s=0.16h^{-1}$. Thus,
for example, in the case $\mu_s=0.04h^{-1}$ the contrast of $\mu_t$
(the ratio of $\mu_t$ in the target to the background value) varied
from $2.0$ in the letters RADIOL to $4.8$ in the letters DEPT.  In the
case $\mu_s=0.16h^{-1}$, the contrast was smaller and varied from
$1.18$ to $2.12$. We note that the contrast in the total attenuation
coefficient depends weakly on the background absorption coefficient,
compared to its dependence on the background scattering coefficient.

The sources were normally incident on the surface
$z=0$. The detectors were placed on the opposite side of the sample
and the specific intensity exiting the surface $z=L_z$ at the angle of
$\pi/4$ with respect to the $z$-axis was measured. In this situation
there are two possibilities---the exiting rays either make an angle of
$\pi/4$ or $3\pi/4$ with the $y$-axis. In some cases, data from both
directions were used. Note that the distance $L_z$ corresponds to the
slab thickness $L$ used in Sec.~\ref{sec:math}. The optical depth of
the sample, $\mu_s L_z$, varied from $1.6$, for $\mu_{s}=0.04h^{-1}$,
to $6.4$, for $\mu_{s}=0.16h^{-1}$ .  This corresponds to the
mesoscopic scattering regime in which the image reconstruction method
of SSOT is applicable.

Reconstruction of the total attenuation coefficient $\mu_t$ was
performed in slices $x=x_{\rm slice}$ separated by a distance $\Delta
x=h$. For each slice, the source positions were $x=x_{\rm slice}$,
$y=nh$, $z=0$, with $n$ being integers. The reconstruction area inside
each slice was $44h\leq y \leq 77h$, $4h\leq z \leq 37h$, with the
field of view $34h \times 34h$. At the noise levels $n=0$ and $n=1\%$,
only the rays making an angle of $\pi/4$ with the $y$-axis were used;
for the noise level $n=3\%$, the exiting rays which make an angle of
$3\pi/4$ with the $y$-axis were also used in order to improve image
quality of the reconstructions. Also, the regularization parameter
$\epsilon$ in the regularized pseudoinverse (\ref{L_star_L_inv}) was
varied to obtain the best visual appearance of images. Note that the
absolute values of the reconstructed $\mu_{t}$ are not sensitive to
the choice of $\epsilon$.  Qualitatively, the same results are
obtained by setting $\epsilon=0$, although we have found that
selecting a small but nonzero value of $\epsilon$ tends to slightly
improve image quality.

The results of image reconstructions for various noise levels are
presented in Figs.~\ref{mu_s=0.04}--\ref{mu_s=0.16}, where we show
both slices containing inhomogeneities and neighboring slices in which
inhomogeneities are not present. It can be seen that the spatial
resolution depends on the noise level and contrast, and can be as good
as one discretization step, $h$.  Note that the reconstructed images
are in very good quantitative agreement with the model (all panels in
each figure are plotted using the same color scale) and stable in the
presence of noise. When $\mu_s=0.16h^{-1}$ (Fig.~\ref{mu_s=0.16}) the
optical depth of the sample is $\mu_s L_z=6.4$. This is a borderline
case when scattering is sufficiently strong so that the
single-scattering approximation of SSOT may be expected to be
inaccurate. Indeed, the image quality in Fig.~\ref{mu_s=0.16} is
markedly worse than in Figs.~\ref{mu_s=0.04} and \ref{mu_s=0.08}, yet
the letters in the image remain legible.
           
\begin{figure}
\begin{center}
\includegraphics[height=19cm]{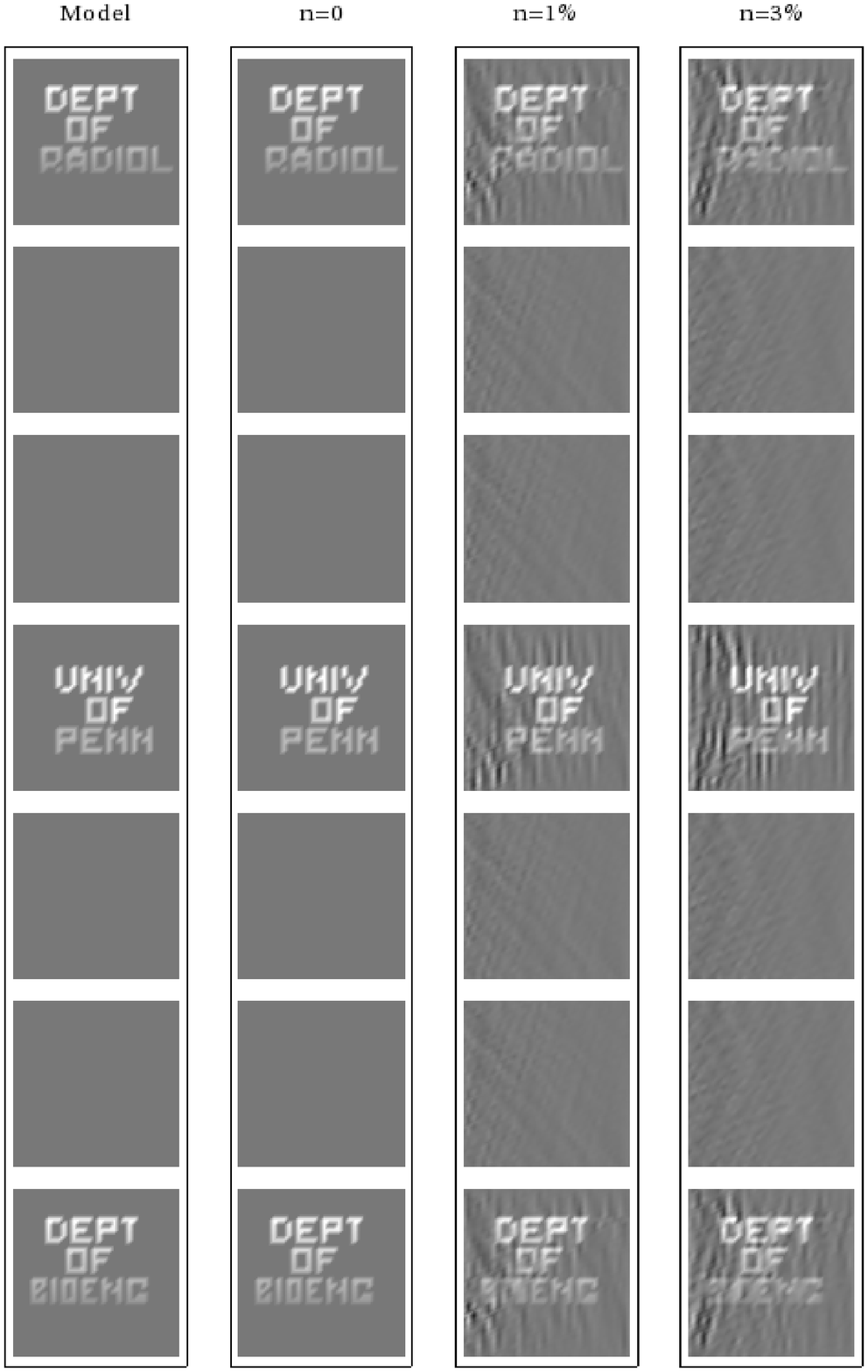}
\end{center}
\caption{\label{inv_crime_mu_s=0.16}(Color online) Image
  reconstruction for $\mu_s=0.16h^{-1}$, for a data function
  corresponding only to single-scattered light and calculated
  according to (\ref{SSOT_2D}).  All the other details are as for
  Fig.~\ref{mu_s=0.16}.}
\end{figure}

We emphasize that the reconstructed images presented here are based on
simulated data obtained by solving the RTE exactly, thereby accounting
for {\em all} orders of scattering. For samples which are optically
thick, the resulting reconstructions evidently exhibit artifacts due
to the breakdown of the single scattering approximation (which is not
possible physically but achievable in simulations). Such would be
absent if only single-scattered light were detected. To illustrate
this idea, we present in Fig.~\ref{inv_crime_mu_s=0.16} reconstructed
images for the case $\mu_s =0.16h^{-1}$ using forward data in which
only single-scattered light is retained.  Here, instead of solving the
full RTE, the data function was directly calculated from
(\ref{Ray_int_3}). The aforementioned procedure of generating data
overestimates the performances of the imaging method. It can be seen from Fig.~8 that, as
expected, significantly better reconstructions result.

\section{Discussion}
\label{sec:conclusions}
We have investigated the problem of optical tomography in the
mesoscopic regime. Within the accuracy of the single scattering
approximation to the RTE, we have derived a relation between the
absorption and scattering coefficients and the specific intensity. In
particular, for a homogeneously scattering medium, we have shown that
the intensity measured by an angularly-selective detector is related
to the integral of the attenuation coefficient along a broken ray.  By
inverting this relation, we are able to recover the attenuation
coefficient of the medium.

The image reconstruction technique we have implemented breaks down in
the multiple scattering regime. In future work, we plan to explore
corrections to the single scattering approximation. In addition, since
electromagnetic waves in random media are, in general, polarized, we
also intend to explore the effects of polarization within the
framework of the generalized vector radiative transport
equation~\cite{ishimaru}. Finally, we note that a technique for
fluorescence imaging of mesoscopic objects has recently been
reported~\cite{ntziachristos}. It would thus be of interest to
investigate the fluorescent analog of SSOT.

\section*{Acknowledgment}
We are grateful to Guillaume Bal and Alexander Katsevich for valuable
discussions. This work was supported by the NSF under the grant
DMS-0554100 and by the NIH under the grant R01EB004832.

 \appendix*
\section{Derivation of Eq.~(\ref{I_s_2_1})}
\label{app:int}

Substitution of (\ref{I_b}) into (\ref{I_s}) with ${\bf
  r}={\bf r}_2$ and $\hat{\bf s}=\hat{\bf s}_2$ results in the
following expression for the single-scattered intensity:

\begin{eqnarray}
I_s({\bf r}_1,\hat{\bf s}_1;{\bf r}_2,\hat{\bf s}_2) = I_0 A(\hat{\bf s}_2,\hat{\bf s}_1) \int \mu_s({\bf r}) g({\bf r}_2,{\bf r})
g({\bf r},{\bf r}_1) \nonumber \\ 
\hspace{1cm} \times \delta (\hat{\bf u}({\bf r}_2 - {\bf r}) - \hat{\bf s}_2)
\delta (\hat{\bf u}({\bf r} - {\bf r}_1) - \hat{\bf s}_1) d^3 r \ ,
\label{app:1}
\end{eqnarray}

\noindent
where we have introduced the notation $\hat{\bf u}(\Br)=\Br/r$. In the
following analysis, the manipulation of delta functions is done in
accordance with the theory of generalized functions
\cite{Lighthill_book}.

We now make the change of variables ${\bf r} = {\bf r}_1 + {\bf R}$, ${\bf
  R}=R\hat{\bf R}$ and $d^3r = d^3R = R^2dR d^2\hat{R}$. The integral
over $d^2\hat{R}$ is immediately evaluated and 
(\ref{app:1}) becomes

\begin{eqnarray}
I_s({\bf r}_1,\hat{\bf s}_1;{\bf r}_2,\hat{\bf s}_2) = I_0 A(\hat{\bf s}_2,\hat{\bf s}_1) \int g({\bf r}_2,{\bf r}_1+R\hat{\bf s}_1)
g({\bf r}_1+R\hat{\bf s}_1,{\bf r}_1) \nonumber \\ 
\hspace{1cm} \times \mu_s({\bf r}_1+R\hat{\bf s}_1) 
\delta (\hat{\bf u}({\bf r}_{21} - R\hat{\bf s}_1) - \hat{\bf s}_2)
R^2 dR \ .
\label{app:2}
\end{eqnarray}

\noindent
We then write the remaining delta-function as

\begin{equation}
\label{app:3}
\delta (\hat{\bf u}-\hat{\bf s}_2) = \delta(\varphi_{\hat{\bf u}} -
\varphi_{\hat{\bf s}_2}) \delta\left( \cos\theta_{\hat{\bf u}} -
  \cos\theta_{\hat{\bf s}_2} \right) \ .
\end{equation}

\noindent
Here ${\bf u}={\bf r}_{21} - R\hat{\bf s}_1$ and $\theta$ and
$\varphi$ are polar angles of the respective unit vectors. It is convenient to work in a reference frame whose $z$-axis coincides with the
source-detector line.  We then find that $\varphi_{\hat{\bf u}} =
\varphi_{\hat{\bf s}_1} \pm \pi$. Consequently,

\begin{equation}
\label{app:4}
\delta(\varphi_{\hat{\bf u}} - \varphi_{\hat{\bf s}_2}) = \delta\left(
  \vert \varphi_{\hat{\bf s}_1} - \varphi_{\hat{\bf s}_2} \vert - \pi
\right) \ .
\end{equation}

\noindent
We next write

\begin{equation}
\label{app:5}
\delta\left(\cos\theta_{\hat{\bf u}} - \cos\theta_{\hat{\bf s}_2}
\right) = \delta(f(R)) \ , 
\end{equation}

\noindent
where 

\begin{equation}
\label{app:6}
f(R)= \frac{r_{21} - R\cos\theta_1}{\sqrt{r_{21}^2 -
    2r_{21}R\cos\theta_1 + R^2}} - \cos\theta_2 \ .
\end{equation}

\noindent
It can be verified that if $\theta_1 + \theta_2\geq \pi$, the equation
$f(R)=0$ has no positive roots. In the opposite limit, however, there
is one positive root $R=L_1$. Note that the lengths $L_1$ and $L_2$ are defined by
(\ref{L_1_2_def}) and illustrated in
Fig.~\ref{fig:sketch2}. We thus have

\begin{equation}
\label{app:7}
R^2 \delta(f(R)) = \Theta(\pi- \theta_1 - \theta_2) L_1^2
\frac{\delta(R-L_1)}{\vert f^{\prime}(L_1) \vert} \ .
\end{equation}

\noindent 
Computation of the above derivative is straightforward and yields

\begin{equation}
\label{app:8}
\vert f^{\prime}(L_1) \vert = \frac{L_1}{r_{21} L_2 \sin^2(\theta_1 +
  \theta_2)} \ .
\end{equation}

\noindent
Collecting everything, we arrive at

\begin{eqnarray}
R^2 \delta(\hat{\bf u}({\bf r}_{21} &-& R\hat{\bf s}_1) - \hat{\bf
  s}_2) = \nonumber \\
&&\delta\left(
  \vert \varphi_{\hat{\bf s}_1} - \varphi_{\hat{\bf s}_2} \vert - \pi
\right) \Theta(\pi- \theta_1 - \theta_2) \frac{r_{21}L_1 L_2
  \delta(R-L_1)}{\sin^2(\theta_1 + \theta_2)} \ .
\label{app:9}
\end{eqnarray}

\noindent
We then recall that ${\bf r}_1 + L_1\hat{\bf s} = {\bf R}_{21}$,
$L_1=\vert {\bf R}_{21} - {\bf r}_1 \vert$, $L_2=\vert {\bf r}_2 -
{\bf R}_{21} \vert$ and obtain

\begin{eqnarray}
\label{app:10_1}
g({\bf r}_2,{\bf r}_1+L_1\hat{\bf s}_1)& =& \frac{1}{L_2^2} \exp\left[ -
  \int_0^{L_2} \mu_t({\bf R}_{21} + \hat{\bf s}_2 \ell)d\ell \right ]
\ , \\
\label{app:10_2}
g({\bf r}_1+L_1\hat{\bf s}_1,{\bf r}_1)& =& \frac{1}{L_1^2} \exp\left[ -
  \int_0^{L_1} \mu_t({\bf r}_1 + \hat{\bf s}_1 \ell)d\ell \right ] \ .
\end{eqnarray}

\noindent
Finally, using (\ref{L_1_2_def}),  we arrive at the
result~(\ref{I_s_2_1}).


\begin{thebibliography}{99}
\bibitem{condition-number} The measure of how numerically
  well-conditioned the problem associated with solving the linear
  equation $Ax=b$ is the condition number of the matrix $A$. This can
  be regarded as the rate at which the solution, $x$, changes with
  respect to a change in $b$, and can be calculated as the ratio
  between the maximal and minimal singular values of $A$,
  respectively. A problem with a low condition number is said to be
  well-conditioned, while a problem with a high condition number is
  said to be ill-conditioned.
\end{thebibliography}
\end{document}